\documentclass[11pt,twoside]{article}

\usepackage{graphicx}
\usepackage{asp2004}
\usepackage{epsf}
\usepackage{psfig}
\usepackage{lscape}

\def\degr{\mathrm{^\circ}}
\def\micron{\,\mu\mathrm{m}}

\def\aap{A\&A}
\def\aaps{A\&AS}
\def\apj{ApJ}
\def\apjl{ApJL}

\def\mnras{MNRAS}

\begin{document}
\title{The vicinity of the galactic supergiant B[e] star CPD$-57\mathrm{^\circ}\,2874$ from near- and mid-IR long baseline
spectro-interferometry with the VLTI (AMBER and MIDI)}   

\author{A.~Domiciano~de~Souza$^1$, T.~Driebe$^1$, O.~Chesneau$^2$, K.-H.~Hofmann$^1$,
S.~Kraus$^1$, A.~S.~Miroshnichenko$^{1,3}$, K.~Ohnaka$^1$,
R.~G.~Petrov$^4$, Th.~Preibisch$^1$, P.~Stee$^2$, G.~Weigelt$^1$}

\affil{$^1$ Max-Planck-Institut f\"{u}r Radioastronomie, Auf dem
H\"{u}gel 69, 53121 Bonn, Germany}    

\affil{$^2$ Observatoire de la C{\^o}te d'Azur, Gemini, CNRS UMR 6203,
Avenue Copernic, 06130 Grasse, France}    

\affil{$^3$ Dept. of Physics and Astronomy, P.O. Box 26170,
University of North Carolina at Greensboro, Greensboro, NC 27402--6170, USA}    

\affil{$^4$ Laboratoire Universitaire d'Astrophysique de Nice
(LUAN), CNRS UMR 6525, UNSA, Parc Valrose, 06108 Nice, France}    

\begin{abstract} 
We present the first spectro-interferometric observations of the
circumstellar envelope (CSE) of a B[e] supergiant
(CPD$-57\mathrm{^\circ}\,2874$), performed with the Very Large
Telescope Interferometer (VLTI) using the beam-combiner
instruments AMBER (near-IR interferometry with three 8.3~m Unit
Telescopes or UTs) and MIDI (mid-IR interferometry with two UTs).
Our observations of the CSE are well fitted by an elliptical
Gaussian model with FWHM diameters varying linearly with
wavelength. Typical diameters measured are
$\simeq1.8\times3.4$~mas or $\simeq4.5\times8.5$~AU (adopting a
distance of $2.5$~kpc) at $2.2\micron$, and $\simeq12\times15$~mas
or $\simeq30\times38$~AU at $12\micron$. We show that a spherical
dust model reproduces the SED but it underestimates the MIDI
visibilities, suggesting that a dense equatorial disk is required
to account for the compact dust-emitting region observed.
Moreover, the derived major-axis position angle in the mid-IR
($\simeq144\degr$) agrees well with previous polarimetric data,
hinting that the hot-dust emission originates in a disk-like
structure. Our results support the non-spherical CSE paradigm for
B[e] supergiants.
\end{abstract}



\section{Introduction}

CPD$-57\mathrm{^\circ}\,2874$ is a poorly-studied object, for
which McGregor et al. (1988) suggested a distance of $d=2.5$ kpc,
assuming that it belongs to the Carina OB association. A high
reddening and the presence of CO emission bands at $2.3-2.4
\micron$ makes it compatible with the supergiant B[e] (sgB[e])
class. Zickgraf (2003) obtained high-resolution optical spectra
exhibiting double-peaked emission lines that are suggestive of a
flattened CSE geometry, typical for sgB[e] stars. However, the
physical parameters of neither the star nor its CSE have been
studied in detail yet.

\section{Results}
\label{results}

We report here our results on CPD$-57\mathrm{^\circ}\,2874$ based
on recent (Dec/2004 - Feb/2005) spectro-interferometric
observations performed with the VLTI instruments AMBER (e.g.,
Petrov et al. 2003) and MIDI (e.g., Leinert et al. 2004).

\begin{figure}[t]
 \centering
  \includegraphics*[width=13cm,draft=false]{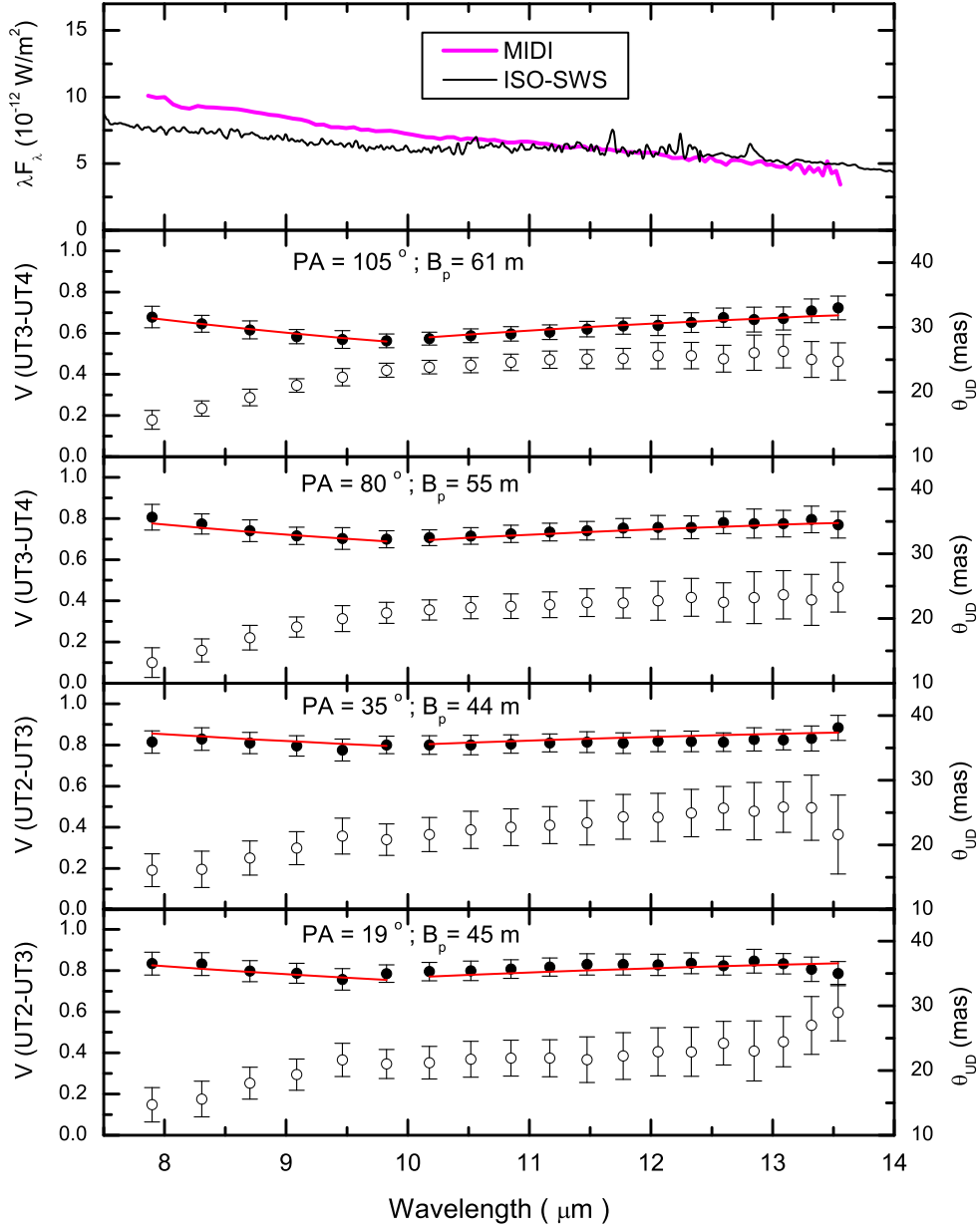}
  \caption{VLTI/MIDI observations of CPD$-57\mathrm{^\circ}\,2874$ obtained in the mid-IR with
spectral resolution $R=30$. $V$ and $\theta_\mathrm{UD}$ are shown
as filled and open circles, respectively. The MIDI and ISO-SWS
(Sloan et al. 2003) spectra (top panel) show no clear evidence of
a silicate feature around $10\micron$. The MIDI visibilities are
well fitted with an elliptical Gaussian model (solid lines;
Eqs.~\ref{eq:gaussian_ellipse_model_V} and \ref{eq:model_size},
and Table~\ref{ta:models}).}
  \label{fig:MIDI_results}
%
\end{figure}

\begin{figure}[t]
 \centering
  \includegraphics*[width=6.0cm,draft=false]{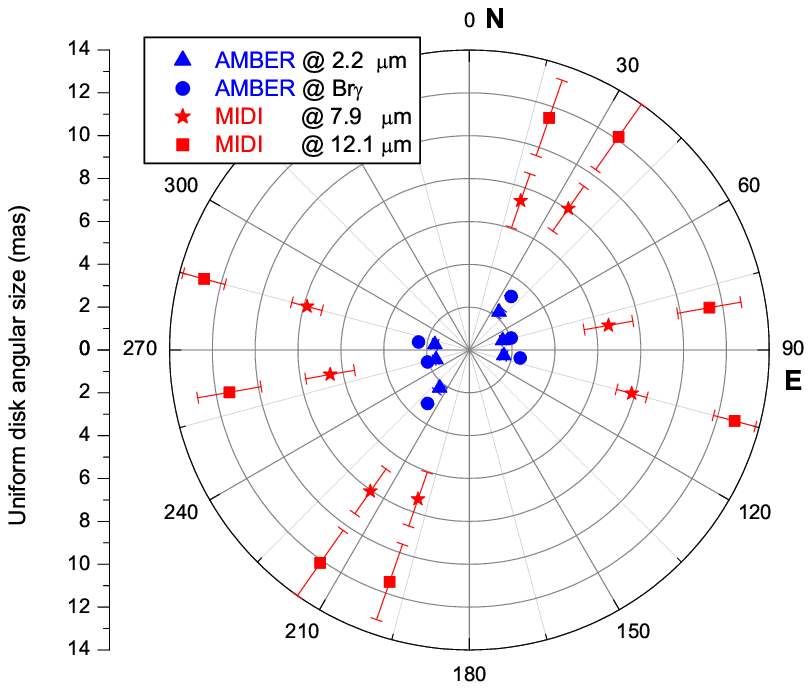}
  \includegraphics*[width=7.0cm,draft=false]{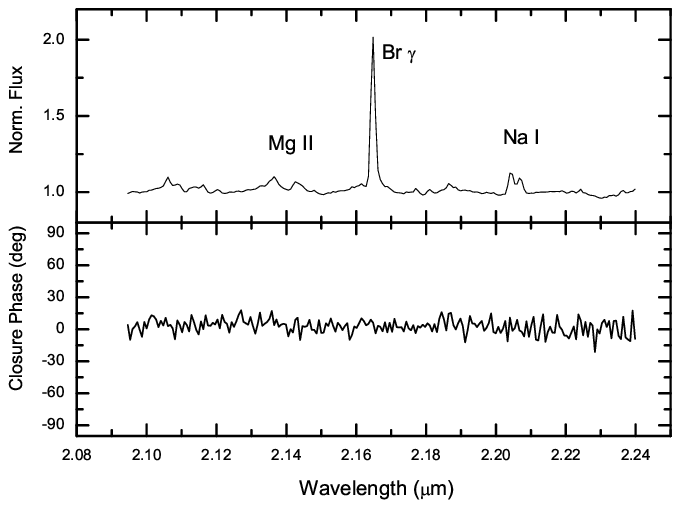}
  \caption{\textit{Left:} Uniform disk sizes
dependence on the baseline position angle (in degrees) for 2
selected wavelengths both from AMBER (continuum at $2.2\micron$
and center of the Br$\gamma$ line) and from MIDI ($7.9\micron$ and
$12.1\micron$). The central star is supposed to be at the center
of the plot and the distance between 2 symmetrical points gives
the corresponding uniform disk angular diameter
($\theta_\mathrm{UD}$) of the CSE. The error bars on the AMBER
$\theta_\mathrm{UD}$ have similar sizes as their corresponding
symbols. Although the $\theta_\mathrm{UD}$ is a zero-order
estimate, this figure allows us to directly compare the CSE sizes
in the near- and mid-IR. \textit{Right:} VLTI/AMBER closure phase
for CPD$-57\mathrm{^\circ}\,2874$ obtained around Br$\gamma$ with
$R=1500$. Within the noise level ($\simeq \pm 10\degr$) the
closure phase is zero (continuum and Br$\gamma$ line) suggesting a
centrally-symmetric intensity distribution in the near-IR.
Spherical and non-spherical (with axial-symmetry) CSEs are thus
compatible with the zero closure phase measured with AMBER.}
  \label{fig:UD_polar_plots}
%
\end{figure}

Figures~\ref{fig:AMBER_results} and \ref{fig:MIDI_results} show
the spectra and visibilities obtained with AMBER and MIDI,
respectively. CPD$-57\mathrm{^\circ}\,2874$ is resolved in both
spectral regions at all projected baselines $B_\mathrm{p}$ and
position angles PA. As a zero-order size estimate these figures
also show the uniform disk angular diameters $\theta_\mathrm{UD}$
obtained from the visibilities at each spectral channel.
Figure~\ref{fig:UD_polar_plots} (left) shows the
$\theta_\mathrm{UD}$ as a function of the baseline position angle
observed with AMBER and MIDI. The measured sizes are clearly
different on the 4 spectral channels chosen ($2.2\micron$,
Br$\gamma$ line, $7.9\micron$, and $12.1\micron$); in particular,
the mid-IR sizes are much larger than those in the near-IR.
Indications of a flattened CSE is more evident in the near-IR than
in the mid-IR but, as we will show in Sect.~\ref{discussion}, both
AMBER and MIDI observations suggest a non-spherical CSE.

The AMBER observations show a zero closure phase
(Fig.~\ref{fig:UD_polar_plots} right) at all wavelengths (within
the noise level of a few degrees). This is a strong indication
that the near-IR emitting regions (continuum and Br$\gamma$ line)
have an approximately centrally-symmetric intensity distribution.

Since sgB[e] stars are thought to have non-spherical winds, we
expect an elongated shape for their CSE, unless the star is seen
close to pole-on. Hereafter, we show that both AMBER and MIDI
observations can indeed be well reproduced by an elliptical
Gaussian model for the CSE intensity distribution, corresponding
to visibilities of the form:
\begin{equation}\label{eq:gaussian_ellipse_model_V}
    V(u,v)=   \exp{ \left\{ \frac{-\pi^2(2a)^2}{4\ln2}[u^2+(Dv)^2] \right\} }
\end{equation}
where $u$ and $v$ are the spatial-frequency coordinates, $2a$ is
the major-axis FWHM of the intensity distribution (image plane),
and $D$ is the ratio between the minor and major axes FWHM
($D=2b/2a$). Since, in general, $2a$ forms an angle $\alpha$ with
the North direction (towards the East), $u$ and $v$ should be
replaced in Eq.~\ref{eq:gaussian_ellipse_model_V} by
($u\sin\alpha+v\cos\alpha$) and ($u\cos\alpha-v\sin\alpha$),
respectively. A preliminary analysis of $V$ at each individual
$\lambda$ showed that $D$ and $\alpha$ can be considered
independent on $\lambda$ within a given spectral band ($K$ or
$N$). On the other hand, the size varies with $\lambda$, as one
can see from the $\theta_\mathrm{UD}(\lambda)$ curves in
Figs.~\ref{fig:AMBER_results} to \ref{fig:UD_polar_plots}.

\subsection{Size and geometry in the $K$ band}

\begin{figure}[t]
 \centering
  \includegraphics*[width=13cm,draft=false]{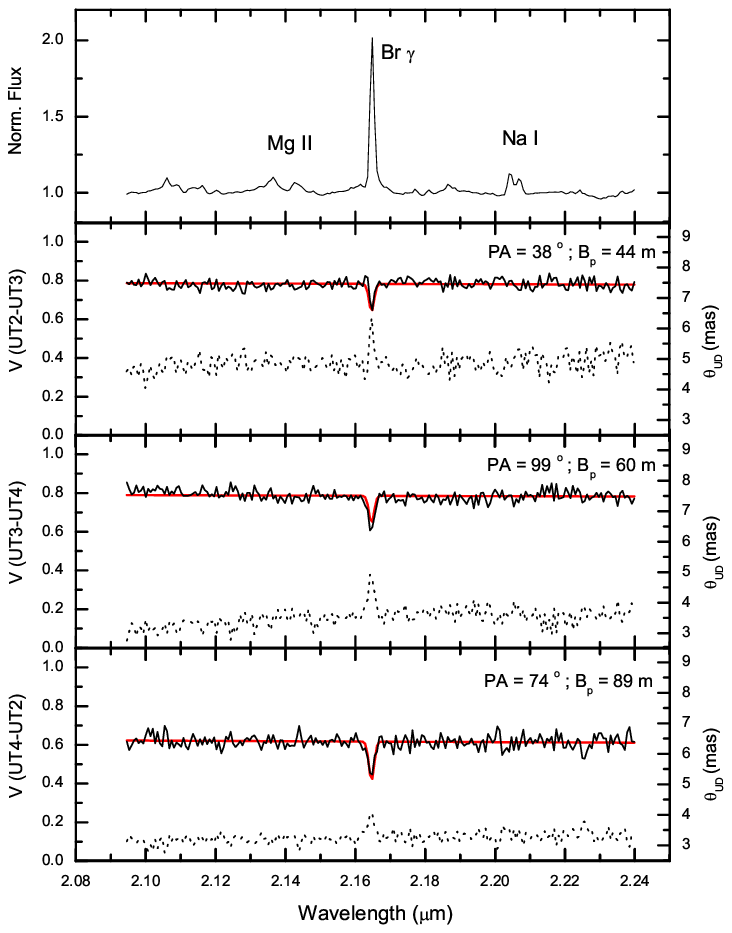}
  \caption{VLTI/AMBER observations of CPD$-57\mathrm{^\circ}\,2874$ obtained
around Br$\gamma$ with spectral resolution $R=1500$. The
normalized flux is shown in the top panel and the visibilities $V$
for each baseline are given in the other panels (the corresponding
projected baselines $B_\mathrm{p}$ and position angles PA are
indicated). The errors in $V$ are $\simeq \pm 5\%$. The dotted
lines are the uniform disk angular diameters $\theta_\mathrm{UD}$
in mas (to be read from the scales on the right axis), computed
from $V$ at each $\lambda$ as a zero-order size estimate. The
visibilities obtained from the elliptical Gaussian model fits the
observations quite well (smooth solid lines;
Eqs.~\ref{eq:gaussian_ellipse_model_V} and \ref{eq:model_size},
and Table~\ref{ta:models}). In contrast to the Br$\gamma$ line,
the Mg~{\sc ii} and Na~{\sc i} lines do not show any clear
signature on the visibilities.}
  \label{fig:AMBER_results}
%
\end{figure}

We interpret the AMBER observations in terms of an elliptical
Gaussian model (Eq.~\ref{eq:gaussian_ellipse_model_V}) with a
chromatic variation of the size. The $\theta_\mathrm{UD}(\lambda)$
curves in Fig.~\ref{fig:AMBER_results} suggest a linear increase
of the size within this part of the $K$ band. In addition, the
AMBER visibilities decrease significantly inside Br$\gamma$,
indicating that the line-forming region is more extended than the
region responsible for the underlying continuum. Based on these
considerations, we adopted the following expression for the
major-axis FWHM:
\begin{equation}\label{eq:model_size}
    2a(\lambda)=2a_0 + C_1(\lambda-\lambda_0) +
C_2 \exp{ \!\!\left[
{\!-4\ln2\left(\frac{\lambda-\lambda_{\mathrm{Br}\gamma}}{\Delta
\lambda}\right)^2} \right] }
\end{equation}
where $2a_0$ is the major-axis FWHM at a chosen reference
wavelength $\lambda_0(=2.2\micron)$, and $C_1$ is the slope of
$2a(\lambda)$. The size increase within Br$\gamma$ is modelled by
a Gaussian with an amplitude $C_2$ and FWHM $\Delta\lambda$,
centered at $\lambda_{\mathrm{Br}\gamma}=2.165\micron$.
Figure~\ref{fig:AMBER_results} shows a rather good fit of this
model to the observed visibilities both in the continuum and
inside Br$\gamma$. The parameters derived from the fit are listed
in Table~\ref{ta:models}.

%

%

\subsection{Size and geometry in the $N$ band}

Similarly to the analysis of the AMBER visibilities, we interpret
the MIDI observations of CPD$-57\mathrm{^\circ}\,2874$ in terms of
an elliptical Gaussian model
(Eq.~\ref{eq:gaussian_ellipse_model_V}) with a size varying
linearly with $\lambda$ as given in Eq.~\ref{eq:model_size} (for
the analysis of the MIDI data the parameter $C_2$ is set to zero).
Additionally, since $\theta_\mathrm{UD}$ shows a stronger
$\lambda$-dependence between 7.9 and 9.8 $\micron$ compared to the
region between 10.2 and 13.5 $\micron$ (see
Fig.~\ref{fig:MIDI_results}), we performed an independent fit for
each of these two spectral regions. This elliptical Gaussian model
provides a good fit to the MIDI visibilities as also shown in
Fig.~\ref{fig:MIDI_results}. The parameters corresponding to the
fit in the two spectral regions within the $N$ band are listed in
Table~\ref{ta:models}.

\begin{table*}[t]
 \centering
 \begin{minipage}[]{\hsize}
   \renewcommand{\footnoterule}{}
 \centering
\caption[]{Model parameters and $\chi^2_\mathrm{red}$ (reduced
chi-squared) derived from the fit of an elliptical Gaussian
(Eqs.~\ref{eq:gaussian_ellipse_model_V} and \ref{eq:model_size})
to the VLTI/AMBER and VLTI/MIDI visibilities. Angular sizes (in
mas) correspond to FWHM diameters. The errors on the fit
parameters are dominated by the calibration errors of the
instrumental transfer function, derived from the calibrator stars.
}\label{ta:models} \vspace{0.1cm}
\begin{tabular}{lccc}
\hline\noalign{\smallskip}
Parameter & AMBER & MIDI ($<10\micron$) & MIDI ($>10\micron$) \\
\hline\noalign{\smallskip}

$\lambda_0$ ($\micron$) & 2.2 & 8.0 & 12.0  \\

major axis $2\,a_0$ (mas) & $3.4\pm0.2$ & $10.1 \pm 0.7$ & $15.3 \pm $ 0.7\\

$C_1$   (mas/$\micron$)  & $ 1.99\pm0.24$ &  $2.58 \pm 0.41$ & $0.45 \pm 0.22$   \\

position angle $\alpha$ ($\degr$)   &$\!\!\!\!\!173\degr\pm9\degr$ & $145\degr \pm 6\degr$  & $143\degr \pm 6\degr$   \\

$D=2b/2a$   & $ 0.53\pm0.03$ & $0.76 \pm 0.11$  & $0.80 \pm 0.10$   \\

minor axis  $2\,b_0$ (mas) & $1.8\pm0.1$ & $7.7 \pm 1.0$  & $12.2 \pm 1.1$   \\

\textbf{Br$\gamma$:} $C_2$ (mas)  & $1.2\pm0.1$ & -- & -- \\

 \textbf{Br$\gamma$:} $\Delta\lambda$  ($10^{-3}\micron$)  & $1.8\pm0.2$ & -- & --  \\

$\chi^2_\mathrm{red}$   & 0.7 & 0.1 & 0.1 \\

\noalign{\smallskip}\hline\noalign{\smallskip}
\end{tabular}
\end{minipage}
\end{table*}
%

\section{Discussion and conclusions} \label{discussion}

\begin{figure}[ht]
 \centering
  \includegraphics*[width=70mm,angle=-90]{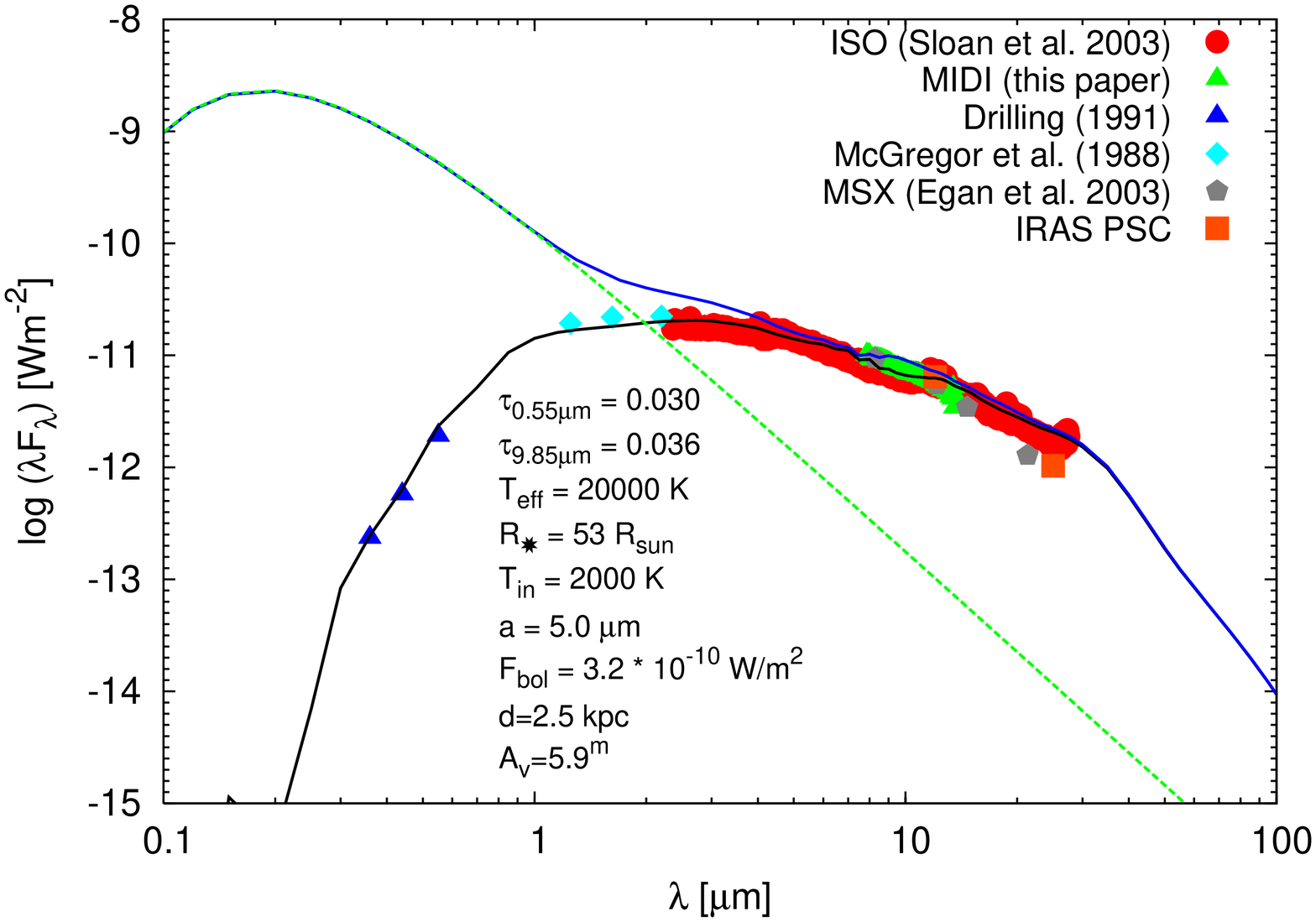}
  \caption{The SED of CPD$-57\mathrm{^\circ}\,2874$
constructed from several sources (see figure label). The solid
lines are the observed and dereddened SEDs modelled with the code
DUSTY (Ivezi\'c \& Elitzur 1997), while the dashed line indicates
the assumed blackbody spectrum of the central star. The parameters
adopted for this model are listed in the figure: $T_{\rm in}$ is
the temperature at the inner dust shell boundary, $a$ is the
diameter of the silicate grains, and the other symbols have their
usual meanings. A standard $r^{-2}$ dust-density law was assumed.}
  \label{fig:DUSTY_silicates}
%
\end{figure}

To further investigate the geometry of the CSE, we attempted to
simultaneously fit the MIDI visibilities and the spectral energy
distribution (SED) using the spherical 1D code DUSTY (Ivezi\'c \&
Elitzur 1997). To reproduce the featureless spectrum around
$10\micron$, we used large silicate grains and/or carbonaceous
dust. The nature of the dust and the featureless mid-IR spectrum
will be addressed in detail elsewhere. This spherical model in the
optically-thin regime can nicely fit the SED (as shown in
Fig.~\ref{fig:DUSTY_silicates}), but it significantly
underestimates the mid-IR visibilities, even when the model
parameters are varied considerably. This means that the measured
dust-emitting region is too compact ($2a\sim 10-16~\mathrm{mas} $
or $ 25-40~\mathrm{AU}$; see Table~\ref{ta:models}) to be
reproduced by a spherical model (even though the SED fit is
acceptable). Only a disk-like structure seems to allow the dust to
survive within $\sim10$~AU from the star (see Kraus \& Lamers
2003).

Another argument against a spherical CSE comes from polarization
measurements of Yudin \& Evans (1998). After correction for the
interstellar polarization, Yudin (private communication) estimated
an intrinsic polarization position angle $\simeq45\degr-55\degr$.
Interestingly, within the error bars this angle is perpendicular
to the major-axis PA we derived from the MIDI data
($\alpha\simeq144\degr$; see Table~\ref{ta:models}), as is
expected from a disk-like dusty CSE.

Thanks to the unprecedent combination of interferometric
resolution, multi-spectral wavelength coverage and relatively high
spectral resolution now available from the VLTI we measured the
size and geometry of the CSE of a sgB[e] star in the near- and
mid-IR. We hope that the present work will open the door for new
spectro-interferometric observations of these complex and
intriguing objects as well as motivate the development of
interferometry-oriented and physically-consistent models for such
objects. A more complete description of the present work is given
by Domiciano de Souza et al.~(2005).

\vspace*{4mm}
\acknowledgements 

A.D.S.\ acknowledges the Max-Planck-Institut f\"{u}r
Radio\-astro\-nomie for a postdoctoral fellowship. We are indebted
to Dr.\ R.\ V.\ Yudin for his calculations on the intrinsic
polarization vector.


\end{document}